\documentclass[sigconf]{acmart}

\def\ba{{\bf a}}
\def\bb{{\bf b}}

\def\bk{{\bf k}}

\def\bn{{\bf n}}
\def\bs{{\bf s}}

\def\bx{{\bf x}}
\def\by{{\bf y}}
\def\bz{{\bf z}}

\def\0{{\bf 0}}
\def\1{{\bf 1}}

\def\bC{{\bf C}}
\def\bD{{\bf D}}
\def\bE{{\bf E}}
\def\bG{{\bf G}}

\def\bI{{\bf I}}

\def\bS{{\bf S}}

\def\bK{{\bf K}}

\def\cL{{\mathcal L}}
\def\cN{{\mathcal N}}

\def\mbR{{\mathbb R}}

\def\etal{\emph{et al. }}
\def\ie{\emph{i.e. }}

\def\grad{{\nabla}}

\usepackage{multirow}
\usepackage{tabularx}
\usepackage{graphicx}
\usepackage{overpic}
\usepackage{amsmath}
\usepackage{flushend}

\def\m{\mathbf}

\DeclareMathOperator{\prox}{Prox}

\def\BibTeX{{\rm B\kern-.05em{\sc i\kern-.025em b}\kern-.08emT\kern-.1667em\lower.7ex\hbox{E}\kern-.125emX}}

\copyrightyear{2019} 
\acmYear{2019} 
\acmConference[ICDSC 2019]{13th International Conference on Distributed Smart Cameras}{Sept. 9--11, 2019}{Trento, Italy}
\acmBooktitle{13th International Conference on Distributed Smart Cameras (ICDSC 2019), Sept. 9--11, 2019, Trento, Italy}
\acmPrice{15.00}
\acmDOI{10.1145/3349801.3349823}
\acmISBN{978-1-4503-7189-6/19/09}
\begin{document}

%
\title{Deep Super-Resolution Network for Single Image Super-Resolution with Realistic Degradations}

%

\author{Rao Muhammad Umer}
\affiliation{%
  \institution{University of Udine}
  \city{Udine}
  \country{Italy.}}
\email{engr.raoumer943@gmail.com}

\author{Gian Luca  Foresti}
\affiliation{%
  \institution{University of Udine}
  \city{Udine}
  \country{Italy.}}
\email{gianluca.foresti@uniud.it}

\author{Christian Micheloni}
\affiliation{%
  \institution{University of Udine}
  \city{Udine}
  \country{Italy.}}
\email{christian.micheloni@uniud.it}

%

%
\begin{abstract}
  Single Image Super-Resolution (SISR) aims to generate a high-resolution (HR) image of a given low-resolution (LR) image. The most of existing convolutional neural network (CNN) based SISR methods usually take an assumption that a LR image is only bicubicly down-sampled version of an HR image. However, the true degradation (i.e. the LR image is a bicubicly downsampled, blurred and noisy version of an HR image) of a LR image goes beyond the widely used bicubic assumption, which makes the SISR problem highly ill-posed nature of inverse problems. To address this issue, we propose a deep SISR network that works for blur kernels of different sizes, and different noise levels in an unified residual CNN-based denoiser network, which significantly improves a practical CNN-based super-resolver for real applications. Extensive experimental results on synthetic LR datasets and real images demonstrate that our proposed method not only can produce better results on more realistic degradation but also computational efficient to practical SISR applications.     
\end{abstract}

%

%

%
\maketitle

\section{Introduction}
The goal of the single image super-resolution (SISR) is to restore high-resolution (HR) image from its low-resolution (LR) counterpart. SISR problem is a classical problem with various practical applications~\cite{Yue2016ImageST} in  satellite imaging, medical imaging, astronomy, microscopy imaging, seismology, remote sensing, surveillance, biometric, etc. In the surveillance field and in particular in case distributed cameras networks~\cite{jcatel2014}, the possibility to transfer low resolution images is a very important feature that allows to share like visual content for detection~\cite{GLetal2003}, classification~\cite{Rani2015}, analysis~\cite{Garcia2016} and network management~\cite{Dieber2011}. SISR methods can be classified into three main categories, \ie, interpolation-based methods, model-based optimization methods, and discriminative learning methods. Interpolation-based methods \ie nearest-neighbor, bilinear and bicubic interpolators are efficient and simple, but have very limited reconstruction image quality. Model-based optimization methods such as non-local self-similarity prior~\cite{Dong2013NonlocallyCS,Lefkimmiatis2017NonlocalCI,Mairal2009NonlocalSM}, sparsity prior~\cite{Yang2010ImageSV} and denoiser prior~\cite{bigdeli2017deep,Egiazarian2015SingleIS,kai2017ircnncvpr}, 
have powerful image priors to reconstruct HR images, but their optimization procedure is computationally expensive. Model-based optimization methods with integration of deep CNN priors can improve efficiency, but due to hand-designed parameters, they are not suitable for end-to-end deep learning.
On the other hand, discriminative learning methods have attracted significant attentions due to
their effectiveness and efficiency for SISR performance by using deep convolution neural networks.\\
The most widely-used degradation model, which is known as bicubic degradation is given as:
\begin{equation}\label{eq:eq_degradation1}
  \mathbf{y} = \mathbf{x}\downarrow_s,
\end{equation}
where the LR image $\by$ is degraded bicubicly from a clean HR image. But, this simple degradation gives inferior results in many practical super-resolution applications.\\
The another more realistic degradation model ~\cite{kai2018srmdcvpr}, in which the LR image $\by\in \mbR^N$ is mathematically described as a blur kernel $\bk\in \mbR^{N \times N}$ convolved with the latent sharp image $\bx\in\mbR^N$. The subsequent downsampling operation is applied on the blurred image and further degraded by an additive noise. This degradation process is given as follows:
\begin{equation}\label{eq:eq_degradation2}
  \textbf{\emph{y}} = (\textbf{\emph{k}} * \textbf{\emph{x}})\downarrow_s + ~ \textbf{\emph{n}},
\end{equation}
where $*$ denotes the convolution operator, $\downarrow_s$ is a down-sampling operator with scale factor $s$, and $\bn\in \mbR^m$ denotes an i.i.d. additive white Gaussian noise (AWGN) term with unknown standard deviation $\sigma$ (i.e. noise level). Equation~\eqref{eq:eq_degradation2} refers to as a general degradation model for SISR. The common blur kernel $\bk$ choice is isotropic or anisotropic Gaussian blur kernel by standard deviation with fixed kernel width \cite{kai2018srmdcvpr}. The more realistic case used in deblurring task is motion blur kernel with arbitrary sizes. Since the LR images also contain noise, where the simple case is to take assumption of AWGN with non-blind noise levels $\sigma$, but more complex scenario is to consider AWGN with blind noise levels $\sigma$. The most popular choice is to use bicubic downsampler operator in SISR methods. Due to unknown noise level and the loss of high-frequency information, which makes the SISR is an highly ill-posed nature of inverse problem, and therefore it is an active and challenging research topic in low-level image processing, computer vision, mobile vision, and computational photography.
The contribution of this paper are as follows:
\begin{itemize}
  \vspace{0.1cm}
  \item We  follow more realistic degradation model than simple bicubic degradation model for SISR, which also considers blur kernels of arbitrary sizes, and different noise levels to take the advantage of existing deblurring methods for blur kernel estimation and denoising.
  \vspace{0.1cm}
  \item A deep single image super-resolution network is proposed to solve SISR with the modified degradation model~\eqref{eq:eq_degradation3}, which goes beyond bicubic degradation and can restore HR image from LR images with different blur kernels.
  \vspace{0.1cm}
  \item The proposed SRWDNet is well designed as the iterative strategy aims to solve the degradation model by minimization of energy function, which makes useful step towards practical applications.
\end{itemize}

\section{Related Work}
The preliminary CNN-based method to solve SISR is super-resolution convolutional network (SRCNN)~\cite{dong2014srcnneccv} network, where a three layer super-resolution network was proposed. In the extension of SRCNN~\cite{dong2016srcnntpami} work, the authors showed the impact of depth of super-resolution network during training a deep neural network, which limits the performance of CNN-based super-resolvers. To address this training difficulty, Kim~\etal~\cite{kim2016vdsrcvpr} proposed a very deep super-resolution (VDSR) network with residual learning approach. To improve the efficiency, the efficient sub-pixel convolutional network (ESPCNN)~\cite{Shi2016pixelcnncvpr} was proposed to take bicubicly LR input and introduced an efficient sub-pixel convolution layer to upscale the LR feature maps to HR images at the end of the network. While achieving the good performance, the above methods take the LR input image as bicubicly downsampled version of HR image, those not only suffer from high computational cost but also hinder the efficiency of practical super-resolution applications due to mismatch of image degradation models.\\
Beyond the widely-used bicubic degradation in the above CNN-based methods, there is an interesting approach of CNN-based methods to solve SISR problem by using model-based optimization frameworks~\cite{chen2017tnrdtpami, Lefkimmiatis2017NonlocalCI, Lefkimmiatis2018UniversalDN, kai2017ircnncvpr}. Besides that, an accurate estimate of blur kernel plays a vital role than sophisticated image priors, pointed in \cite{Efrat2013AccurateBM}. Since then, several methods have been proposed to tackle LR images that go beyond bicubic degradation to solve the energy function induced by equation~\eqref{eq:eq_degradation2}. Zhang~\etal proposed iterative residual convolutional network (IRCNN)~\cite{kai2017ircnncvpr} to solve SISR problem by using a plug-and-play framework. Zhang~\etal proposed a deep CNN-based super-resolution with multiple degradation (SRMD)~\cite{kai2018srmdcvpr}, which takes two degradation parameters (\ie blur kernel $\bk$, and $\sigma$) as input to the network, but they only consider Gaussian blur kernels with fixed kernel width. \\
The above SISR methods have three main drawbacks. First, they have difficulty in complex (e.g. motion) blur kernel estimation with arbitrary dimensions. Second, they are usually designed for Gaussian blur kernels with fixed kernel dimension and thus cannot tackle severely blurred LR image effectively. Third, they have not trained a unified network, which handle blur kernel estimation, noise levels, and scaling factor within a single network by training end-to-end fashion.

\begin{figure*}[t]
\centering
\begin{minipage}[b]{1\textwidth}
\centerline{
\begin{overpic}[trim=0 0 0 0, clip, width=1\textwidth]
{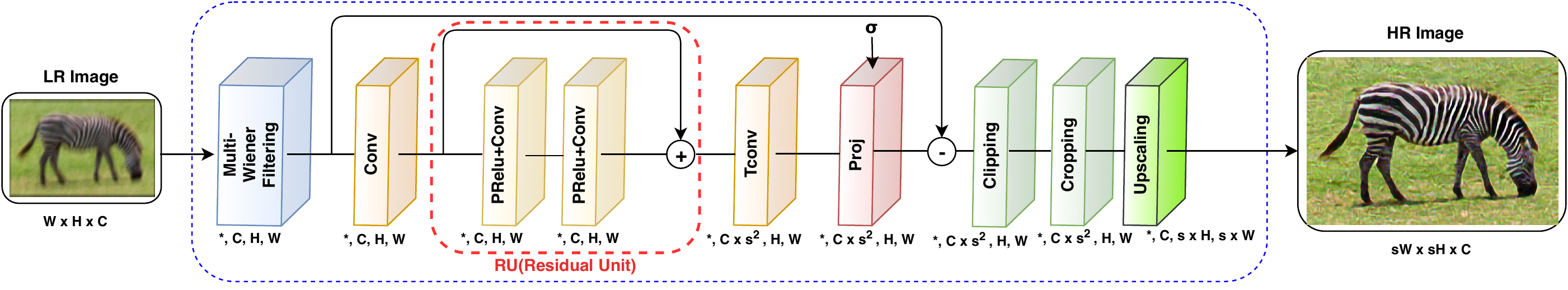}
\end{overpic}}
\end{minipage}
\vspace{-0.4cm}
\caption{SRWDNet architecture. Our network takes an input LR (low-resolution) image, blur kernel $\bk$, noise sigma $\mathbf{\sigma}$, and up-scaling factor $\bs$, then reconstructs an HR (high-resolution) image of the given scaling factor $\bs$. The LR image is $W\times H \times C$ dimension, while HR image is $\bs W\times \bs H \times C$, where $C$ is the number of channels of the input image, and $\bs$ is the upscaling factor.  
}
\label{fig:srwdnet}
\end{figure*}

\section{Problem Formulation}
The degradation model proposed in ~\cite{kai2018srmdcvpr} is given as:
\begin{equation}\label{eq:eq_degradation3}
  \textbf{\emph{y}}= \textbf{\emph{k}} * (\textbf{\emph{x}}\downarrow_s)  + ~ \textbf{\emph{n}},
\end{equation}
where $\downarrow_s$ is the bicubic downsampler with scale factor $s$. Equation ~\eqref{eq:eq_degradation3} corresponds to a deblurring problem followed by a SISR problem with general degradation. This model has distinctive advantage over equation~\eqref{eq:eq_degradation2} as it estimates the blur kernel efficiently for existing deblurring methods and also holds the degradation assumption of equation~\eqref{eq:eq_degradation1}.\\
After finalizing the suitable degradation model, we formally define the energy function according to Maximum A Posteriori (MAP) framework by referencing the equation~\eqref{eq:eq_degradation3}, and  given as follows:
\begin{equation}\label{eq:eng_func} 
\hat{\bx} = \arg\underset{x}{\min} ~\frac{1}{2\sigma^2}\|\by - \bk*(\bx\downarrow_s)\|_2^2 + \lambda \varphi(\bx),
\end{equation}
where $\frac{1}{2\sigma^2}\|\by - \bk*(\bx\downarrow_s)\|_2^2$ is the data fidelity (log-likelihood) term that quantifies the proximity of the solution to the observations, while $\varphi(\bx)$ is regularization term associated with image prior, $\sigma$ is the unknown noise level i.e. belongs to AWGN noise, and $\lambda$ is the trade-off parameter (i.e. governs the compromise between noise reduction and details preservation). The DNN-based inference models usually correspond to an energy function for discriminative learning, where the degradation model is defined by the training LR and HR pairs. It demonstrates that why existing DNN-based SISR trained network on bicubic degradation (refers to ~\eqref{eq:eq_degradation1}) has limited performance for real super-resolution applications.

\subsection{Optimization Strategy}
In this section, we briefly give the overview of optimization strategy for our network training. By referring to equation~\eqref{eq:eng_func}, we want to recover the underlying image $\bx$ as the minimizer of the objective function as: 
\begin{equation} 
\hat{\bx} = \arg\underset{x}{\min} ~\bE(\bx),
\label{eq:e1}
\end{equation}
As the energy function $\bE(.)$ consists of data fidelity term and regularizer term, which is given as:
\begin{equation} 
\hat{\bx} = \arg\underset{x}{\min} ~\bD(\bx;\bk, \by, \downarrow_{\bs}) + \lambda \varphi(\bx),
\label{eq:e2}
\end{equation}
So, overall objective function~\eqref{eq:eng_func} can be formally rewritten as a constrained optimization form: 
\begin{equation} 
\hat{\bx} = \underbrace{\arg\min_{a\leqslant x \leqslant b} ~\frac{1}{2\sigma^2}\|\by - \bk*(\bx\downarrow_s)\|_2^2 + \lambda \varphi(\bx)}_{\mathbf{f}(\bx)},
\label{eq:e3}
\end{equation}
To solve the Eq.~\eqref{eq:e3}, there are several modern convex-optimization schemes for large-scale problems such as Split-Bregman~\cite{split_bregman}, HQS method~\cite{HQS}, ADMM ~\cite{boyd_admm}, Primal-dual algorithms~\cite{Chambolle2011Primaldual}, and Proximal methods~\cite{ParikhPGM}. In our work, we solve the Eq.~\eqref{eq:e3} by using the Proximal Gradient Method (PGM)~\cite{ParikhPGM}, which is a generalization of gradient descent algorithm. PGM~\cite{ParikhPGM} deals with the optimization of a function that is not fully differentiable, but it can be split into a smooth and a non-smooth part. To do so, we first rewrite Eq.~\eqref{eq:e3} as:
\begin{equation} 
\hat{\bx} = \arg\underset{x}{\min} ~\mathbf{f}(\bx) + \mathbf{i}_{c}(\bx),
\label{eq:e4}
\end{equation}
where $\mathbf{i_c}$ is the indicator function of the convex set $\bC \in \{ \bx \in  \mbR^m : \ba \leqslant \bx_k \leqslant \bb, \forall k \}$.
The gradient of $\mathbf{f}(\bx)$ is computed as:
\begin{equation}
    \nabla_{\m x}\mathbf{f}(\bx) = \frac{1}{\sigma^2}\bK^{T}(\bK(\bx\downarrow_s) - \by) + \lambda \Psi(\bx),
    \label{eq:e5}
\end{equation}
Then, the solution of Eq.~\eqref{eq:e4} is computed in an iterative fashion by using following update:
\begin{equation}
    \bx_t\downarrow_s = \prox_{\gamma^{t}\mathbf{i_c}}\left(\bx_{(t-1)}\downarrow_s - \gamma^t\nabla_{\m x}\mathbf{f}(\bx_{(t-1)})\right),
    \label{eq:e6}
\end{equation}
where $\gamma^t$ is a step-size and $\prox_{\gamma^{t}\mathbf{i_c}}$ is the proximal operator~\cite{ParikhPGM} related to the indicator function $\mathbf{i_c}$, which can be defined as: 
\begin{equation}
     \prox_{h}(\bz) = \arg\underset{x \in \bC}{\min} ~\frac{1}{2}\|\bx - \bz\|_2^2 + h(\bx),
    \label{eq:e7}
\end{equation}
Since proximal map $\prox_{\gamma\sigma^2}$ gives the regularized solution of a Gaussian denoising problem, so finally we have the following form of our solution as: 
\begin{equation}
    \bx_t = \left(\prox_{\gamma^{t}\sigma^2}\left((1-\gamma^t\bK^{T}\bK)(\bx_{(t-1)})\downarrow_s + ~\gamma^t \bK^T\by - ~\lambda\gamma^t \Psi(\bx_{t-1})\right)\right)\uparrow_s,
    \label{eq:e8}
\end{equation}
where $\uparrow_s$ is the upscaling operator. Thus, we design the network by unrolling $\bS$ stages of equation \ref{eq:e8} between the proximal input and the super-resolution output. For the proposed network, the objective function is minimized by discriminative learning as:
\begin{equation} 
    \left\lbrace \begin{array}{l}\arg\underset{\Theta}{\min}~\mathcal{L}(\Theta) = \sum \limits _{s = 1}^{S}\frac{1}{2} \Vert \hat{\bx}_T^s - \bx_{gt}^s\Vert ^2_2\\ \mathrm{s.t.} \left\lbrace \begin{array}{l}\bx_0^s = \bI_0^s \\ update~ \bx_{t}^s ~according~ to~ Eq.~\eqref{eq:e8}, \\ t = 1 \ldots T\, \end{array}\right.\end{array}\right. 
    \label{eq:e9}
\end{equation}
where, $\Theta = \{\Theta\}_{t=1}^{t=T}$, and $\bI_0$ is the initial value of the regularizer term. It can be noted that the above loss function only depends upon the final iteration $\mathbf{T}$, where the network parameters in all stages $\mathbf{S}$ are optimized simultaneously. This minimization training strategy is usually called \emph{joint training}, similar to~\cite{Lefkimmiatis2018UniversalDN,chen2017tnrdtpami, Schmidt2014ShrinkageFF}.  

\section{Proposed Network}
The proposed network architecture for non-blind SISR is shown in figure~\ref{fig:srwdnet}. The input of our network is LR image $\by$ with the corresponding blur kernel $\bk$, noise sigma $\sigma$, and scaling factor $\bs$. Our network first applies deconvolution operation on the LR blurry and noisy input via deconvolution module, estimate the noise variance by the denoising module, and finally the HR image by the upscaling module.

\subsection{Deconvolution module}
In our proposed network, the deconvolution module is the learnable Wiener Filtering layer as shown in Figure \ref{fig:srwdnet}. In Wiener filtering layer, we formulate the following objective function as:
\begin{equation} 
\hat{\bx} = \arg\underset{x}{\min} ~\frac{1}{2}\|\by - \bK\bx\|_2^2 + \frac{\alpha}{2} \|\bG\bx\|_2^2,
\label{eq:wf1}
\end{equation}
Where $\by\in \mbR^N$ is the observation, $\bK \in \mbR^{N \times N}$ is the blur kernel, and $\bG \in \mbR^{N\times N}$ is the regularization kernel, and both (\ie $\bK$ and $\bG$) are considered as the circulant matrices. In case of multiple regularization kernels, the equation \ref{eq:wf1} can be written as:
\begin{equation} 
\hat{\bx} = \underbrace{\arg\underset{x}{\min} ~\frac{1}{2}\|\by - \bK\bx\|_2^2 + \frac{\alpha}{2}~\sum_{i=1}^{d} \|\bG_i\bx\|_2^2}_{\mathbf{f}(\bx)},
\label{eq:wf2}
\end{equation}
where $G_{i}$ plays the role of multiple regularizer filters, and the closed-form solution of equation \ref{eq:wf2} can be computed by Wiener deconvolution technique \cite{WienerFilter}. So, we learn the Eq.~\eqref{eq:wf2} as following form in Wiener filtering layer:
\begin{equation}\label{eq:wf3}
 \boldsymbol{\hat{\textbf{\emph{x}}}} = \mathcal{F}(\textbf{\emph{y}}, \textbf{\emph{k}},\sigma; \Theta),
\end{equation}
Where $\Theta$ denotes the trainable regularization kernels weights by gradient descent update rule in the network. Here, we compute the gradient of $\mathbf{f}(\bx)$ as:
\begin{equation}
    \nabla_{\m x}\mathbf{f}(\bx) = \bK^{T}(\bK\bx - \by) + \alpha\sum_{i=1}^{d}\bG_{i}^{T}\bG_i\bx
    \label{eq:wf4}
\end{equation}
After re-arranging Eq.~\eqref{eq:wf4}, we have the following closed-form solution as: 
\begin{equation}
    \label{eq:min_eng_grad}
    \hat{\bx} = (\bK^{T}\bK + \alpha \sum_{i=1}^{d} \bG_{i}^T\bG_i)^{-1} \bK^T \by,
\end{equation}
where we take $\alpha \leftarrow e^\alpha$ (\ie [0.0001, 0.01]). The weights of Wiener Convolution layer (\ie $\Theta$) are 24 output features map with kernel size $5\times$5 by initializing the discrete cosine transform (DCT) basis, which are updated according to PGM (refers to eq.~\eqref{eq:e8}). 
\subsection{Denoising module}
Since there are many image denoising neural networks such as the DnCNN \cite{Zhang2017BeyondAG}, IRCNN \cite{kai2017ircnncvpr}, and UDNet \cite{Lefkimmiatis2018UniversalDN}, but we use UDNet \cite{Lefkimmiatis2018UniversalDN} as a residual CNN denoiser, which has less number of trainable parameters and helps to efficiently approximate the proximal map. Since UDNet~\cite{Lefkimmiatis2018UniversalDN} has less trainable parameters, so it can be useful to practical SISR photography applications. The architecture of UDNet~\cite{Lefkimmiatis2018UniversalDN} is consist of $N$ residual units with 2 convolution layers each of 64 kernels by $3\times3$ filter size, and each convolution layer is preceded by the parametrized rectified linear unit (PReLU)~\cite{He2015DelvingDI}. In figure~\ref{fig:srwdnet}, we use five residual unit (RU) blocks, which are sandwich by convolution and transpose convolution layer with shared parameters. Both layers (\ie \emph{Conv} and \emph{TConv}) have 64 features map by $7\times7$ kernel size with $C \times H\times W$ tensors, where $C$ is the number of channels of the input image $\by$. In our proposed network, the denoiser module can be replaced by the other CNN-based denoising networks, which exhibits the similar characteristics like UDNet~\cite{Lefkimmiatis2018UniversalDN}.    

\subsection{Upscaling module}
Finally, an efficient sub-pixel convolution~\cite{Shi2016pixelcnncvpr} layer with a stride of $1/s$ is followed by the last transpose convolutional layer to convert multiple latent images of size $s^2C \times H \times W$ to a single HR image of size $sW \times sH \times C$.

\section{Experimental Setup}
The experimental performance of our proposed network is measured by the peak signal-to-noise ratio (PSNR) and structural similarity (SSIM) measure. In the further sections, we provide you the details of our network training parameters setting, trainset, testset, comparison with others SISR methods, and computational cost of our method.
\subsection{Network training parameters setting}
To train the proposed network, the image patch size is set to $256\times256$ by center cropping the image. We use the ADAM~\cite{Kingma2015AdamAM} optimizer with a single batch size for training with the loss function as described in section~\ref{sec:loss_criteria}. We set the fixed learning rate as 0.001 and the default values of $\beta_1$ and $\beta_2$ (0.9 and 0.999) of the ADAM optimizer are used. We set the weight decay to 0.0001, and also set \emph{amsgrad} flag as true. For all reported results in this paper, we train the network for 50 epochs, and there are number of iterations in each epoch depends on total batches in the data loader.
\subsection{Loss function}
\label{sec:loss_criteria}
The proposed method is expected to restore the sufficient content of clear image $\bx$ and make the recovered image $\mathbf{\hat{x}}$ sharp. In this work, we choose training loss consists of \textit{content loss} and \textit{gradient loss}:
\begin{equation}
\mathcal{L}= \mathcal{L}_c+\mathcal{L}_{grad},
\label{eq:loss_function}
\end{equation}
Where $\mathcal{L}_c$ is mean squared error (MSE) between the ground truth $\bx$ and the estimated $\mathbf{\hat{x}}$:
\begin{equation}
\cL_{\text{c}}(\bx_i, \hat{\bx}_i; \Theta) = \|\hat{\bx}_i-\bx_i\|_2^2,
\label{eq:loss_2}
\end{equation}
And $\mathcal{L}_{grad}$ is to minimize the gradient discrepancy in the training:
\begin{equation}
\cL_{\text{grad}}(\bx_i, \hat{\bx}_i; \Theta) = \|\grad_v\hat{\bx}_i-\grad_v\bx_i\|_2^2 + \|\grad_h\hat{\bx}_i-\grad_h\bx_i\|_2^2,
\label{eq:loss_grad}
\end{equation}
where $\grad_v$ and $\grad_h$ denote the operators calculating the image gradients in the horizontal and vertical directions, respectively. The loss function in \eqref{eq:loss_grad} is expected to help to produce sharp images. 
\subsection{Training dataset}
\label{sec:train_data}
In order to generate downsampled, blurred, and noisy images for training, we use BSDS500 dataset~\cite{Arbelez2011bsdstpami}, and center cropped image patches with a size of $256 \times256$ pixels as clear images. We take training dataset of 400 high resolution ground-truth images from BSDS500~\cite{Arbelez2011bsdstpami}. We generate 10 randomly motion blurred kernels for training and testing according to ~\cite{Boracchi2012ModelingTP}, whose blur kernel size ranges from $11 \times 11$ to $31 \times 31$ pixels. We bicubicly downsample the clear images with scaling factors $\bs$ (\ie $\times 2, \times 3, \times 4 $), then convolve the downsampled images with the motion blur kernels $\bk$ for training (see supplementary material), and also add Gaussian noises with 1\%, 2\%, 3\%, and 5\% noise standard deviation to generate LR image patches. Instead of training a customized model for blur kernels with fixed dimension and non-blind noise levels, we uniformly sample kernel sizes from a set $[11, 13, 15, 17, 19, 21, 23, 27, 29, 31]$ and noise levels from an interval $[1\%, 2\%, 3\%, 5\%]$ \footnote{A LR image $\by$ with  Gaussian noise $\sigma$ is generated by adding noise from $\cN(\0, \sigma^2)$ for image $\bk*(\bx)\downarrow_s$ with $[0, 255]$ intensity range.}, which helps to learn a more versatile model to handle diverse data.


\subsection{Testing dataset}
\label{sec:test_data}
We evaluate the proposed network on well-known SISR benchmark testing datasets, \ie Set5~\cite{Timofte2014aplusaccv}, Set14~\cite{Timofte2014aplusaccv}, and Urban100~\cite{Huang2015SingleIS}, that are independent to the training dataset. We conduct all experiments on these synthetic LR testing datasets, which are generated by bicubicly downsampling the ground-truth (GT) images with scaling factor $\bs$ (\ie $\times 2, \times 3, \times 4 $), then blurring them with 10 generated motion blur kernels (see supplementary material) of size ranges from $11 \times 11$ to $31 \times 31$ pixels, followed by an addition of AWGN noise level $\sigma$, which includes 1\% (\ie 2.55) noise standard deviation. We generate 50 LR images of Set5~\cite{Timofte2014aplusaccv} with 5 HR GTs, 140 LR images of Set14~\cite{Timofte2014aplusaccv} with 14 HR GTs, and 1000 LR images of Urban100~\cite{Huang2015SingleIS} with 100 HR GTs respectively.
\begin{table*}[ht]
\caption{Average PSNR and SSIM results of SISR methods with more realistic degradation (refers to Eq.~\eqref{eq:eq_degradation3}) on testing datasets, i.e. Set5, Set14, and Urban100.}
\centering
\small\addtolength{\tabcolsep}{-3pt}
\begin{tabular}{|c|c|c|c|c|c|c|c|c|c|c|}
\hline
\multirow{2}{*}{Dataset} & \multicolumn{4}{c|}{Degradation Settings}                                                                                                                                                                                        & Bicubic & VDSR(CVPR)~\cite{kim2016vdsrcvpr} & TNRD(TPAMI)~\cite{chen2017tnrdtpami}& IRCNN(CVPR)~\cite{kai2017ircnncvpr} & SRMD(CVPR)~\cite{kai2018srmdcvpr} & SRWDNet(Ours) \\ \cline{2-11} 
                         & \begin{tabular}[c]{@{}c@{}}Scale\\ Factor\end{tabular} & \begin{tabular}[c]{@{}c@{}}Kernel\\ size\end{tabular} & \begin{tabular}[c]{@{}c@{}}Down-\\ sampler\end{tabular} & \begin{tabular}[c]{@{}c@{}}Noise\\ Level\end{tabular} & \multicolumn{6}{c|}{Average PSNR / SSIM}                        \\ \hline
\multirow{3}{*}{Set5}  & \multirow{1}{*}{$\times2$} & 
                            {\begin{tabular}[c]{@{}c@{}}$11\times11$ to \\ $31\times31$\end{tabular}}                                & \multirow{1}{*}{Bicubic}                                
                              & $1\%$ & 19.30 / 0.5070 & 19.24 / 0.4767 & 19.41 / 0.4937 & 19.00 / 0.4545 & 17.94 / 0.4414 & \bf 23.13 / 0.5870
                          \\ \cline{2-11} 
                         & \multirow{1}{*}{$\times3$}                                     & {\begin{tabular}[c]{@{}c@{}}$11\times11$ to \\ $31\times31$\end{tabular}}                                & \multirow{1}{*}{Bicubic}                                
                              & $1\%$ & 17.90 / 0.4668 & 17.86 / 0.4431  & 17.90 / 0.4765 & 17.63 / 0.4171 & 17.40 / 0.4311 & \bf 21.00 / 0.5025
                         \\ \cline{2-11} 
                         & \multirow{1}{*}{$\times4$}                                     & {\begin{tabular}[c]{@{}c@{}}$11\times11$ to \\ $31\times31$\end{tabular}}                                & \multirow{1}{*}{Bicubic}                                
                              & $1\%$ & 17.01 / 0.4496 & 16.97 / 0.4296  & 17.21 / 0.4609 & 16.74 / 0.4053 & 16.72 / 0.4263 & \bf 20.58 / 0.5036
                         \\ \hline
\multirow{3}{*}{Set14}    & \multirow{1}{*}{$\times2$}                                     &                                            {\begin{tabular}[c]{@{}c@{}}$11\times11$ to \\ $31\times31$\end{tabular}}                                & \multirow{1}{*}{Bicubic}                                
                              & $1\%$ & 18.85 / 0.4419 & 18.80 / 0.4147 & 18.99 / 0.4453 & 18.59 / 0.3981 & 17.15 / 0.3772 & \bf 21.28 / 0.5120  
                         \\ \cline{2-11} 
                         & \multirow{1}{*}{$\times3$}                                     & {\begin{tabular}[c]{@{}c@{}}$11\times11$ to \\ $31\times31$\end{tabular}}                                & \multirow{1}{*}{Bicubic}                                
                              & $1\%$ & 17.74 / 0.4127 & 17.70 / 0.3900 & 17.52 / \textbf{0.4726} & 17.49 / 0.3722 & 17.24 / 0.3858 & \textbf{19.25} / 0.4042         
                        \\ \cline{2-11} 
                         & \multirow{1}{*}{$\times4$}                                     & {\begin{tabular}[c]{@{}c@{}}$11\times11$ to \\ $31\times31$\end{tabular}}                                & \multirow{1}{*}{Bicubic}                                
                              & $1\%$ & 16.99 / 0.4012 & 16.97 / 0.3818 & 17.10 / \textbf{0.4509} & 16.75 / 0.3651 & 16.73 / 0.3842 & \textbf{19.10} / 0.4109 
                        \\ \hline
\multirow{3}{*}{Urban100}   & \multirow{1}{*}{$\times2$}                                     &                                          {\begin{tabular}[c]{@{}c@{}}$11\times11$ to \\ $31\times31$\end{tabular}}                                & \multirow{1}{*}{Bicubic}                                
                              & $1\%$ & 17.30 / 0.4007 & 17.25 / 0.3729   & 17.58 / 0.4336  & 17.01 / 0.4235 & 15.23 / 0.3357 & \bf 19.81 / 0.4914  
                        \\ \cline{2-11} 
                         & \multirow{1}{*}{$\times3$}                                     & 
                        {\begin{tabular}[c]{@{}c@{}}$11\times11$ to \\ $31\times31$\end{tabular}}                                & \multirow{1}{*}{Bicublic}                               
                              & $1\%$ & 16.44 / 0.3773 & 16.41 / 0.3539  & 16.45 / 0.4802  & 16.14 / 0.3523   & 15.85 / 0.3538 & \bf 17.98 / 0.3810
                        \\ \cline{2-11} 
                         & \multirow{1}{*}{$\times4$}                                     & {\begin{tabular}[c]{@{}c@{}}$11\times11$ to \\ $31\times31$\end{tabular}}                                & \multirow{1}{*}{Bicubic}                                
                              & $1\%$ & 15.89 / 0.3694 & 15.87 / 0.3491 & 16.23 / \textbf{0.4608}  & 15.95 / 0.3478  & 15.65 / 0.3601  & \textbf{17.65} / 0.3744
                        \\ \hline
\end{tabular}
\label{tab:t1} 
\vspace{-0.1cm}
\end{table*}

\begin{table*}[ht]
\label{table:comp_time}
\begin{center}
\small\addtolength{\tabcolsep}{-3pt}
\caption{Comparison of the computational time of different SISR methods (Unit:seconds).}
\begin{tabular}{|c|c|c|c|c|c|}
\hline
    ~Degradation Scenario~ & ~VDSR~ & ~TNRD~ & ~IRCNN~ & ~SRMD~ & ~SRWDNet(Ours)~ \\ \hline
  {\begin{tabular}[c]{@{}c@{}}image size: $500\times480$, \\ motion blur kernel: $31\times31$, \\ $\sigma$= 1\%, upscaling factor = $\times 4$\end{tabular}} & 1.573  &  19.573  &  30.561 & 0.305 & 0.593  \\ 
\hline
\end{tabular}
\end{center}
\vspace{-0.4cm}
\end{table*}

\begin{figure*}[!htbp]\footnotesize
	\centering
    \hspace{-0.23cm}
    \begin{tabular}{c@{\extracolsep{0.23em}}c@{\extracolsep{0.23em}}c@{\extracolsep{0.23em}}c@{\extracolsep{0.23em}}c}
    \small\addtolength{\tabcolsep}{-4pt}
		{\begin{overpic}[width=0.20\textwidth]{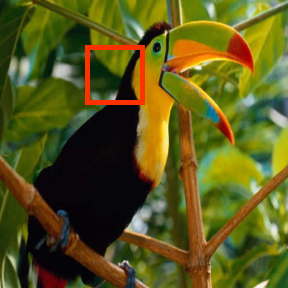}
		\put(70, 0){\includegraphics[width=0.06\textwidth]{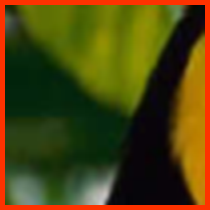}}
		\end{overpic}} 
		&{\begin{overpic}[width=0.20\textwidth]{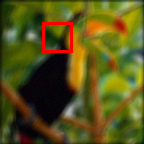}
        \put(75, 75){\includegraphics[width=0.05\textwidth]{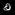}}
        \put(70, 0){\includegraphics[width=0.06\textwidth]{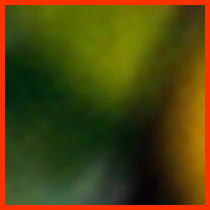}}
        \end{overpic}} 
		&{\begin{overpic}[width=0.20\textwidth]{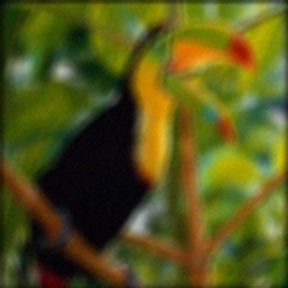}
		\put(70, 0){\includegraphics[width=0.06\textwidth]{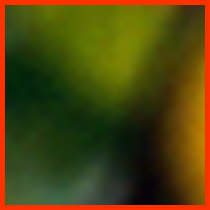}}
		\end{overpic}} 
		&{\begin{overpic}[width=0.20\textwidth]{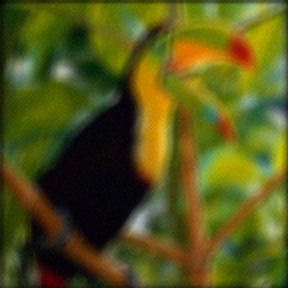}
		\put(70, 0){\includegraphics[width=0.06\textwidth]{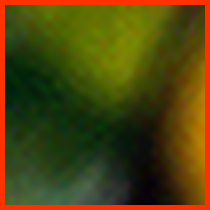}}
		\end{overpic}} \\
        PSNR/SSIM & $\times$2 & (21.33/0.5465) & (21.25/0.5200)\\
		(a) Ground-truth & (b) LR  & (c) Bicubic & (d) VDSR \\
		{\begin{overpic}[width=0.20\textwidth]{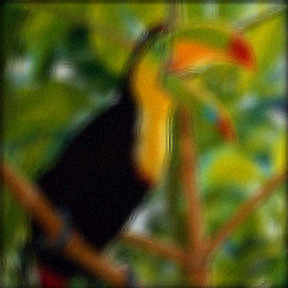}
		\put(70, 0){\includegraphics[width=0.06\textwidth]{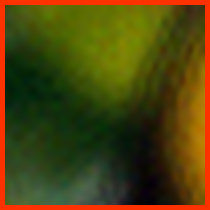}}
		\end{overpic}}
		&{\begin{overpic}[width=0.20\textwidth]{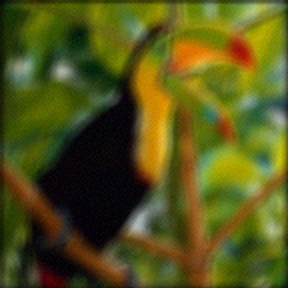}
		\put(70, 0){\includegraphics[width=0.06\textwidth]{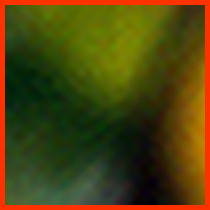}}
		\end{overpic}}
		&{\begin{overpic}[width=0.20\textwidth]{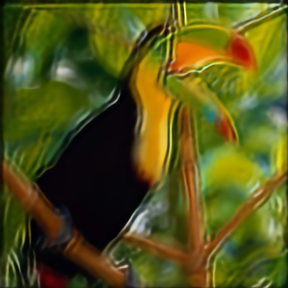}
		\put(70, 0){\includegraphics[width=0.06\textwidth]{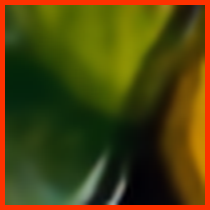}}
		\end{overpic}}
		&{\begin{overpic}[width=0.20\textwidth]{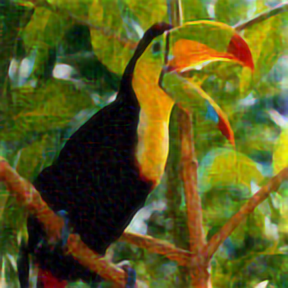}
		\put(70, 0){\includegraphics[width=0.06\textwidth]{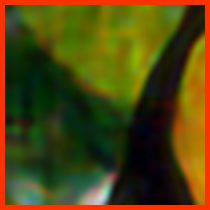}}
		\end{overpic}}\\
		(21.39/0.5323) & (21.23/0.5000) & (19.61/0.4689) & (26.06/0.6817)\\
        (e) TNRD & (f) IRCNN & (g) SRMD & (h) SRWDNet(ours)\\
	\end{tabular}
	\caption{The visual comparison of different SISR methods for scale factor 2 on Set5. The blur kernel is shown on the upper-right corner of the LR image.}
	\label{fig:fig_visual1}
	\vspace{0.1cm}
\end{figure*}
\subsection{Comparisons with state-of-art SISR methods}
We evaluate our proposed SRWDNet on testing SISR benchmark datasets (\ie Set5, Set14, and Urban100) in terms of PSNR and SSIM performance metrics. We compare our proposed method with traditional bicubic method (\ie \emph{imresize} Matlab function used), and other DNN-based SISR methods including VDSR~\cite{kim2016vdsrcvpr}, TNRD~\cite{chen2017tnrdtpami}, IRCNN~\cite{kai2017ircnncvpr}, and SRMD~\cite{kai2018srmdcvpr}. The IRCNN~\cite{kai2017ircnncvpr} and SRMD~\cite{kai2018srmdcvpr} can take degraded image $\by$, blur kernel $\bk$, and noise level $\sigma$ as input, while, VDSR~\cite{kim2016vdsrcvpr} and TNRD~\cite{chen2017tnrdtpami} can take degraded image $\by$ and noise level $\sigma$ as input to the network. For fair comparison, we give the testing image generated according to  degradation model ~\eqref{eq:eq_degradation3} to the above methods.\\
We evaluate our proposed method on SISR testing datasets (\ie Set5, Set14, and Urban100) with different degradation settings and report results in Table 1 in terms of Average PSNR and SSIM. Our method performs well against the others SISR methods. Our method gets much cleaner and HR images with fine texture details without blur and boundary artifacts, while the others methods suffer from over-smoothed images and unpleasant artifacts. Figure \ref{fig:fig_visual1} shows the visual comparison of SISR methods for super-resolving LR image with motion blur kernel by scale factor of $\times$2 (see supplementary material for more results). VDSR produces unpleasant blurred results due to bicubic degradation assumption which deviates from the true one. TNRD also produce unpleasant results due to mismatch of realistic degradation model. Since IRCNN and SRMD follow the true degradation assumption, but SRMD produce more visually pleasant results than IRCNN. SRMD has still blurring artifacts due to opt simple Gaussian blur kernel with fixed width. Even though the input LR image is severely degraded by large downsampling factor, blur kernel and also noisy, our method achieves higher performance both quantitatively and qualitatively than others SISR methods due to obeying the more realistic degradation model. Note that our proposed network not only super-resolved the LR image, but also remove blur and noise from the LR image.
\subsection{Running time}
Our proposed method performs well in terms of computational cost efficiency with other state-of-art SISR methods, which is favorable for practical super-resolution applications. Table 2 shows the testing execution time of respective methods with specific image degradation scenario, measured on our hardware environment\footnote{Hardware environment used: Intel(R) Core(TM) i7-8700 @ 3.20GHz, memory size: 32GB, GPU: Nvidia Quadro P4000}. Testing time of all methods is measured on GPU.   
\subsection{Limitations}
Our method is capable of producing high-quality images from a severely degraded noisy LR images with complex motion blur kernels of arbitrary sizes. However, the main limitation of our network is the unpleasant results when there is a strong presence of noise \ie 3\%, 5\%, or more. Moreover, we train different networks with their respective scaling factors, which limits the performance of our network on other scaling factor for super-resolution. MDSR~\cite{Lim2017EnhancedDR} approach is one possible solution to tackle multiple scaling factors within a same network, but it has not considered the blur kernel and strong noise effect in the LR images. 
\section{Conclusion}
In this paper, we propose an efficient deep SISR network to reconstruct sharp high-resolution images from blurred noisy low-resolution images. The proposed method uses the more realistic degradation model which can benefit existing non-blind deblurring methods for blur kernel estimation. We split the SISR problem into joint deblurring, denoising, and super-resolution tasks and solve it by training the end-to-end network with the proximal gradient descent optimization in an iterative manner. Extensive experimental results show that the proposed method is feasible for the more realistic degradation model and performs favorably against the state-of-art existing methods for SISR in terms of quantitative and visual quality as well as computational cost.
%


\bibliographystyle{ACM-Reference-Format}
\bibliography{refs.bib}


\begin{thebibliography}{36}


\ifx \showCODEN    \undefined \def \showCODEN     #1{\unskip}     \fi
\ifx \showDOI      \undefined \def \showDOI       #1{#1}\fi
\ifx \showISBNx    \undefined \def \showISBNx     #1{\unskip}     \fi
\ifx \showISBNxiii \undefined \def \showISBNxiii  #1{\unskip}     \fi
\ifx \showISSN     \undefined \def \showISSN      #1{\unskip}     \fi
\ifx \showLCCN     \undefined \def \showLCCN      #1{\unskip}     \fi
\ifx \shownote     \undefined \def \shownote      #1{#1}          \fi
\ifx \showarticletitle \undefined \def \showarticletitle #1{#1}   \fi
\ifx \showURL      \undefined \def \showURL       {\relax}        \fi
\providecommand\bibfield[2]{#2}
\providecommand\bibinfo[2]{#2}
\providecommand\natexlab[1]{#1}
\providecommand\showeprint[2][]{arXiv:#2}

\bibitem[\protect\citeauthoryear{Arbelaez, Maire, Fowlkes, and Malik}{Arbelaez
  et~al\mbox{.}}{2011}]%
        {Arbelez2011bsdstpami}
\bibfield{author}{\bibinfo{person}{Pablo~Andres Arbelaez},
  \bibinfo{person}{Michael Maire}, \bibinfo{person}{Charless~C. Fowlkes}, {and}
  \bibinfo{person}{Jitendra Malik}.} \bibinfo{year}{2011}\natexlab{}.
\newblock \showarticletitle{Contour Detection and Hierarchical Image
  Segmentation}.
\newblock \bibinfo{journal}{\emph{IEEE Transactions on Pattern Analysis and
  Machine Intelligence}}  \bibinfo{volume}{33} (\bibinfo{year}{2011}),
  \bibinfo{pages}{898--916}.
\newblock


\bibitem[\protect\citeauthoryear{Bigdeli, Zwicker, Favaro, and Jin}{Bigdeli
  et~al\mbox{.}}{2017}]%
        {bigdeli2017deep}
\bibfield{author}{\bibinfo{person}{Siavash~Arjomand Bigdeli},
  \bibinfo{person}{Matthias Zwicker}, \bibinfo{person}{Paolo Favaro}, {and}
  \bibinfo{person}{Meiguang Jin}.} \bibinfo{year}{2017}\natexlab{}.
\newblock \showarticletitle{Deep mean-shift priors for image restoration}. In
  \bibinfo{booktitle}{\emph{Advances in Neural Information Processing
  Systems}}. \bibinfo{pages}{763--772}.
\newblock


\bibitem[\protect\citeauthoryear{Boracchi and Foi}{Boracchi and Foi}{2012}]%
        {Boracchi2012ModelingTP}
\bibfield{author}{\bibinfo{person}{Giacomo Boracchi} {and}
  \bibinfo{person}{Alessandro Foi}.} \bibinfo{year}{2012}\natexlab{}.
\newblock \showarticletitle{Modeling the Performance of Image Restoration From
  Motion Blur}.
\newblock \bibinfo{journal}{\emph{IEEE Transactions on Image Processing}}
  \bibinfo{volume}{21} (\bibinfo{year}{2012}), \bibinfo{pages}{3502--3517}.
\newblock


\bibitem[\protect\citeauthoryear{Boyd, Parikh, Chu, Peleato, and Eckstein}{Boyd
  et~al\mbox{.}}{2011}]%
        {boyd_admm}
\bibfield{author}{\bibinfo{person}{Stephen Boyd}, \bibinfo{person}{Neal
  Parikh}, \bibinfo{person}{Eric Chu}, \bibinfo{person}{Borja Peleato}, {and}
  \bibinfo{person}{Jonathan Eckstein}.} \bibinfo{year}{2011}\natexlab{}.
\newblock \showarticletitle{Distributed Optimization and Statistical Learning
  via the Alternating Direction Method of Multipliers}.
\newblock \bibinfo{journal}{\emph{Found. Trends Mach. Learn.}}
  \bibinfo{volume}{3}, \bibinfo{number}{1} (\bibinfo{date}{Jan.}
  \bibinfo{year}{2011}), \bibinfo{pages}{1--122}.
\newblock
\showISSN{1935-8237}


\bibitem[\protect\citeauthoryear{Chambolle and Pock}{Chambolle and
  Pock}{2011}]%
        {Chambolle2011Primaldual}
\bibfield{author}{\bibinfo{person}{Antonin Chambolle} {and}
  \bibinfo{person}{Thomas Pock}.} \bibinfo{year}{2011}\natexlab{}.
\newblock \showarticletitle{A First-Order Primal-Dual Algorithm for Convex
  Problems with Applications to Imaging}.
\newblock \bibinfo{journal}{\emph{Journal of Mathematical Imaging and Vision}}
  \bibinfo{volume}{40}, \bibinfo{number}{1} (\bibinfo{date}{01 May}
  \bibinfo{year}{2011}), \bibinfo{pages}{120--145}.
\newblock
\showISSN{1573-7683}


\bibitem[\protect\citeauthoryear{Chen and Pock}{Chen and Pock}{2017}]%
        {chen2017tnrdtpami}
\bibfield{author}{\bibinfo{person}{Yunjin Chen} {and} \bibinfo{person}{Thomas
  Pock}.} \bibinfo{year}{2017}\natexlab{}.
\newblock \showarticletitle{Trainable Nonlinear Reaction Diffusion: A Flexible
  Framework for Fast and Effective Image Restoration}.
\newblock \bibinfo{journal}{\emph{IEEE Transactions on Pattern Analysis and
  Machine Intelligence}}  \bibinfo{volume}{39} (\bibinfo{year}{2017}),
  \bibinfo{pages}{1256--1272}.
\newblock


\bibitem[\protect\citeauthoryear{Dieber, Micheloni, and Rinner}{Dieber
  et~al\mbox{.}}{2011}]%
        {Dieber2011}
\bibfield{author}{\bibinfo{person}{Bernhard Dieber}, \bibinfo{person}{Christian
  Micheloni}, {and} \bibinfo{person}{Bernhard Rinner}.}
  \bibinfo{year}{2011}\natexlab{}.
\newblock \showarticletitle{Resource-Aware Coverage and Task Assignment in
  Visual Sensor Networks}.
\newblock \bibinfo{journal}{\emph{IEEE Trans. Circuits Syst. Video Techn.}}
  \bibinfo{volume}{21} (\bibinfo{date}{10} \bibinfo{year}{2011}),
  \bibinfo{pages}{1424--1437}.
\newblock
\urldef\tempurl%
\url{https://doi.org/10.1109/TCSVT.2011.2162770}
\showDOI{\tempurl}


\bibitem[\protect\citeauthoryear{Dong, Loy, He, and Tang}{Dong
  et~al\mbox{.}}{2014}]%
        {dong2014srcnneccv}
\bibfield{author}{\bibinfo{person}{Chao Dong}, \bibinfo{person}{Chen~Change
  Loy}, \bibinfo{person}{Kaiming He}, {and} \bibinfo{person}{Xiaoou Tang}.}
  \bibinfo{year}{2014}\natexlab{}.
\newblock \showarticletitle{Learning a Deep Convolutional Network for Image
  Super-Resolution}. In \bibinfo{booktitle}{\emph{ECCV}}.
\newblock


\bibitem[\protect\citeauthoryear{Dong, Loy, He, and Tang}{Dong
  et~al\mbox{.}}{2016}]%
        {dong2016srcnntpami}
\bibfield{author}{\bibinfo{person}{Chao Dong}, \bibinfo{person}{Chen~Change
  Loy}, \bibinfo{person}{Kaiming He}, {and} \bibinfo{person}{Xiaoou Tang}.}
  \bibinfo{year}{2016}\natexlab{}.
\newblock \showarticletitle{Image Super-Resolution Using Deep Convolutional
  Networks}.
\newblock \bibinfo{journal}{\emph{IEEE Transactions on Pattern Analysis and
  Machine Intelligence}}  \bibinfo{volume}{38} (\bibinfo{year}{2016}),
  \bibinfo{pages}{295--307}.
\newblock


\bibitem[\protect\citeauthoryear{Dong, Zhang, Shi, and Li}{Dong
  et~al\mbox{.}}{2013}]%
        {Dong2013NonlocallyCS}
\bibfield{author}{\bibinfo{person}{Weisheng Dong}, \bibinfo{person}{Lei Zhang},
  \bibinfo{person}{Guangming Shi}, {and} \bibinfo{person}{Xin Li}.}
  \bibinfo{year}{2013}\natexlab{}.
\newblock \showarticletitle{Nonlocally Centralized Sparse Representation for
  Image Restoration}.
\newblock \bibinfo{journal}{\emph{IEEE Transactions on Image Processing}}
  \bibinfo{volume}{22} (\bibinfo{year}{2013}), \bibinfo{pages}{1620--1630}.
\newblock


\bibitem[\protect\citeauthoryear{Efrat, Glasner, Apartsin, Nadler, and
  Levin}{Efrat et~al\mbox{.}}{2013}]%
        {Efrat2013AccurateBM}
\bibfield{author}{\bibinfo{person}{Netalee Efrat}, \bibinfo{person}{Daniel
  Glasner}, \bibinfo{person}{Alexander Apartsin}, \bibinfo{person}{Boaz
  Nadler}, {and} \bibinfo{person}{Anat Levin}.}
  \bibinfo{year}{2013}\natexlab{}.
\newblock \showarticletitle{Accurate Blur Models vs. Image Priors in Single
  Image Super-resolution}.
\newblock \bibinfo{journal}{\emph{2013 IEEE International Conference on
  Computer Vision}} (\bibinfo{year}{2013}), \bibinfo{pages}{2832--2839}.
\newblock


\bibitem[\protect\citeauthoryear{Egiazarian and Katkovnik}{Egiazarian and
  Katkovnik}{2015}]%
        {Egiazarian2015SingleIS}
\bibfield{author}{\bibinfo{person}{Karen~O. Egiazarian} {and}
  \bibinfo{person}{Vladimir Katkovnik}.} \bibinfo{year}{2015}\natexlab{}.
\newblock \showarticletitle{Single image super-resolution via BM3D sparse
  coding}.
\newblock \bibinfo{journal}{\emph{2015 23rd European Signal Processing
  Conference (EUSIPCO)}} (\bibinfo{year}{2015}), \bibinfo{pages}{2849--2853}.
\newblock


\bibitem[\protect\citeauthoryear{{Foresti}, {Micheloni}, {Snidaro}, and
  {Marchiol}}{{Foresti} et~al\mbox{.}}{2003}]%
        {GLetal2003}
\bibfield{author}{\bibinfo{person}{G.~L. {Foresti}}, \bibinfo{person}{C.
  {Micheloni}}, \bibinfo{person}{L. {Snidaro}}, {and} \bibinfo{person}{C.
  {Marchiol}}.} \bibinfo{year}{2003}\natexlab{}.
\newblock \showarticletitle{Face detection for visual surveillance}. In
  \bibinfo{booktitle}{\emph{12th International Conference on Image Analysis and
  Processing, 2003.Proceedings.}} \bibinfo{pages}{115--120}.
\newblock
\urldef\tempurl%
\url{https://doi.org/10.1109/ICIAP.2003.1234036}
\showDOI{\tempurl}


\bibitem[\protect\citeauthoryear{Garc\'{\i}a, Martinel, Gardel, Bravo, Foresti,
  and Micheloni}{Garc\'{\i}a et~al\mbox{.}}{2016}]%
        {Garcia2016}
\bibfield{author}{\bibinfo{person}{Jorge Garc\'{\i}a}, \bibinfo{person}{Niki
  Martinel}, \bibinfo{person}{Alfredo Gardel}, \bibinfo{person}{Ignacio Bravo},
  \bibinfo{person}{Gian~Luca Foresti}, {and} \bibinfo{person}{Christian
  Micheloni}.} \bibinfo{year}{2016}\natexlab{}.
\newblock \showarticletitle{Modeling Feature Distances by Orientation Driven
  Classifiers for Person Re-identification}.
\newblock \bibinfo{journal}{\emph{J. Vis. Comun. Image Represent.}}
  \bibinfo{volume}{38}, \bibinfo{number}{C} (\bibinfo{date}{July}
  \bibinfo{year}{2016}), \bibinfo{pages}{115--129}.
\newblock
\showISSN{1047-3203}
\urldef\tempurl%
\url{https://doi.org/10.1016/j.jvcir.2016.02.009}
\showDOI{\tempurl}


\bibitem[\protect\citeauthoryear{{Geman} and {Chengda Yang}}{{Geman} and
  {Chengda Yang}}{1995}]%
        {HQS}
\bibfield{author}{\bibinfo{person}{D. {Geman}} {and} \bibinfo{person}{{Chengda
  Yang}}.} \bibinfo{year}{1995}\natexlab{}.
\newblock \showarticletitle{Nonlinear image recovery with half-quadratic
  regularization}.
\newblock \bibinfo{journal}{\emph{IEEE Transactions on Image Processing}}
  \bibinfo{volume}{4}, \bibinfo{number}{7} (\bibinfo{date}{July}
  \bibinfo{year}{1995}), \bibinfo{pages}{932--946}.
\newblock


\bibitem[\protect\citeauthoryear{Goldstein and Osher}{Goldstein and
  Osher}{2009}]%
        {split_bregman}
\bibfield{author}{\bibinfo{person}{T. Goldstein} {and} \bibinfo{person}{S.
  Osher}.} \bibinfo{year}{2009}\natexlab{}.
\newblock \showarticletitle{The Split Bregman Method for L1-Regularized
  Problems}.
\newblock \bibinfo{journal}{\emph{SIAM Journal on Imaging Sciences}}
  \bibinfo{volume}{2}, \bibinfo{number}{2} (\bibinfo{year}{2009}),
  \bibinfo{pages}{323--343}.
\newblock
\urldef\tempurl%
\url{https://doi.org/10.1137/080725891}
\showDOI{\tempurl}


\bibitem[\protect\citeauthoryear{He, Zhang, Ren, and Sun}{He
  et~al\mbox{.}}{2015}]%
        {He2015DelvingDI}
\bibfield{author}{\bibinfo{person}{Kaiming He}, \bibinfo{person}{Xiangyu
  Zhang}, \bibinfo{person}{Shaoqing Ren}, {and} \bibinfo{person}{Jian Sun}.}
  \bibinfo{year}{2015}\natexlab{}.
\newblock \showarticletitle{Delving Deep into Rectifiers: Surpassing
  Human-Level Performance on ImageNet Classification}.
\newblock \bibinfo{journal}{\emph{2015 IEEE International Conference on
  Computer Vision (ICCV)}} (\bibinfo{year}{2015}), \bibinfo{pages}{1026--1034}.
\newblock


\bibitem[\protect\citeauthoryear{Huang, Singh, and Ahuja}{Huang
  et~al\mbox{.}}{2015}]%
        {Huang2015SingleIS}
\bibfield{author}{\bibinfo{person}{Jia-Bin Huang}, \bibinfo{person}{Abhishek
  Singh}, {and} \bibinfo{person}{Narendra Ahuja}.}
  \bibinfo{year}{2015}\natexlab{}.
\newblock \showarticletitle{Single image super-resolution from transformed
  self-exemplars}.
\newblock \bibinfo{journal}{\emph{2015 IEEE Conference on Computer Vision and
  Pattern Recognition (CVPR)}} (\bibinfo{year}{2015}),
  \bibinfo{pages}{5197--5206}.
\newblock


\bibitem[\protect\citeauthoryear{Kim, Lee, and Lee}{Kim et~al\mbox{.}}{2016}]%
        {kim2016vdsrcvpr}
\bibfield{author}{\bibinfo{person}{Jiwon Kim}, \bibinfo{person}{Jung~Kwon Lee},
  {and} \bibinfo{person}{Kyoung~Mu Lee}.} \bibinfo{year}{2016}\natexlab{}.
\newblock \showarticletitle{Accurate Image Super-Resolution Using Very Deep
  Convolutional Networks}.
\newblock \bibinfo{journal}{\emph{2016 IEEE Conference on Computer Vision and
  Pattern Recognition (CVPR)}} (\bibinfo{year}{2016}),
  \bibinfo{pages}{1646--1654}.
\newblock


\bibitem[\protect\citeauthoryear{Kingma and Ba}{Kingma and Ba}{2015}]%
        {Kingma2015AdamAM}
\bibfield{author}{\bibinfo{person}{Diederik~P. Kingma} {and}
  \bibinfo{person}{Jimmy Ba}.} \bibinfo{year}{2015}\natexlab{}.
\newblock \showarticletitle{Adam: A Method for Stochastic Optimization}.
\newblock \bibinfo{journal}{\emph{CoRR}}  \bibinfo{volume}{abs/1412.6980}
  (\bibinfo{year}{2015}).
\newblock


\bibitem[\protect\citeauthoryear{Lefkimmiatis}{Lefkimmiatis}{2017}]%
        {Lefkimmiatis2017NonlocalCI}
\bibfield{author}{\bibinfo{person}{Stamatios Lefkimmiatis}.}
  \bibinfo{year}{2017}\natexlab{}.
\newblock \showarticletitle{Non-local Color Image Denoising with Convolutional
  Neural Networks}.
\newblock \bibinfo{journal}{\emph{2017 IEEE Conference on Computer Vision and
  Pattern Recognition (CVPR)}} (\bibinfo{year}{2017}),
  \bibinfo{pages}{5882--5891}.
\newblock


\bibitem[\protect\citeauthoryear{Lefkimmiatis}{Lefkimmiatis}{2018}]%
        {Lefkimmiatis2018UniversalDN}
\bibfield{author}{\bibinfo{person}{Stamatios Lefkimmiatis}.}
  \bibinfo{year}{2018}\natexlab{}.
\newblock \showarticletitle{Universal Denoising Networks: A Novel CNN
  Architecture for Image Denoising}.
\newblock \bibinfo{journal}{\emph{2018 IEEE/CVF Conference on Computer Vision
  and Pattern Recognition}} (\bibinfo{year}{2018}),
  \bibinfo{pages}{3204--3213}.
\newblock


\bibitem[\protect\citeauthoryear{Lim, Son, Kim, Nah, and Lee}{Lim
  et~al\mbox{.}}{2017}]%
        {Lim2017EnhancedDR}
\bibfield{author}{\bibinfo{person}{Bee Lim}, \bibinfo{person}{Sanghyun Son},
  \bibinfo{person}{Heewon Kim}, \bibinfo{person}{Seungjun Nah}, {and}
  \bibinfo{person}{Kyoung~Mu Lee}.} \bibinfo{year}{2017}\natexlab{}.
\newblock \showarticletitle{Enhanced Deep Residual Networks for Single Image
  Super-Resolution}.
\newblock \bibinfo{journal}{\emph{2017 IEEE Conference on Computer Vision and
  Pattern Recognition Workshops (CVPRW)}} (\bibinfo{year}{2017}),
  \bibinfo{pages}{1132--1140}.
\newblock


\bibitem[\protect\citeauthoryear{Mairal, Bach, Ponce, Sapiro, and
  Zisserman}{Mairal et~al\mbox{.}}{2009}]%
        {Mairal2009NonlocalSM}
\bibfield{author}{\bibinfo{person}{Julien Mairal}, \bibinfo{person}{Francis~R.
  Bach}, \bibinfo{person}{Jean Ponce}, \bibinfo{person}{Guillermo Sapiro},
  {and} \bibinfo{person}{Andrew Zisserman}.} \bibinfo{year}{2009}\natexlab{}.
\newblock \showarticletitle{Non-local sparse models for image restoration}.
\newblock \bibinfo{journal}{\emph{2009 IEEE 12th International Conference on
  Computer Vision}} (\bibinfo{year}{2009}), \bibinfo{pages}{2272--2279}.
\newblock


\bibitem[\protect\citeauthoryear{Parikh and Boyd}{Parikh and Boyd}{2014}]%
        {ParikhPGM}
\bibfield{author}{\bibinfo{person}{Neal Parikh} {and} \bibinfo{person}{Stephen
  Boyd}.} \bibinfo{year}{2014}\natexlab{}.
\newblock \showarticletitle{Proximal Algorithms}.
\newblock \bibinfo{journal}{\emph{Found. Trends Optim.}} \bibinfo{volume}{1},
  \bibinfo{number}{3} (\bibinfo{date}{Jan.} \bibinfo{year}{2014}),
  \bibinfo{pages}{127--239}.
\newblock
\showISSN{2167-3888}


\bibitem[\protect\citeauthoryear{Rani, Foresti, and Micheloni}{Rani
  et~al\mbox{.}}{2015}]%
        {Rani2015}
\bibfield{author}{\bibinfo{person}{Asha Rani}, \bibinfo{person}{Gian~Luca
  Foresti}, {and} \bibinfo{person}{Christian Micheloni}.}
  \bibinfo{year}{2015}\natexlab{}.
\newblock \showarticletitle{A Neural Tree for Classification Using Convex
  Objective Function}.
\newblock \bibinfo{journal}{\emph{Pattern Recogn. Lett.}} \bibinfo{volume}{68},
  \bibinfo{number}{P1} (\bibinfo{date}{Dec.} \bibinfo{year}{2015}),
  \bibinfo{pages}{41--47}.
\newblock
\showISSN{0167-8655}
\urldef\tempurl%
\url{https://doi.org/10.1016/j.patrec.2015.08.017}
\showDOI{\tempurl}


\bibitem[\protect\citeauthoryear{SanMiguel, Micheloni, Shoop, Foresti, and
  Cavallaro}{SanMiguel et~al\mbox{.}}{2014}]%
        {jcatel2014}
\bibfield{author}{\bibinfo{person}{J.~C. SanMiguel}, \bibinfo{person}{C.
  Micheloni}, \bibinfo{person}{K. Shoop}, \bibinfo{person}{G. Foresti}, {and}
  \bibinfo{person}{A. Cavallaro}.} \bibinfo{year}{2014}\natexlab{}.
\newblock \showarticletitle{Self-Reconfigurable Smart Camera Networks}.
\newblock \bibinfo{journal}{\emph{Computer}} \bibinfo{volume}{47},
  \bibinfo{number}{05} (\bibinfo{date}{May} \bibinfo{year}{2014}),
  \bibinfo{pages}{67--73}.
\newblock
\showISSN{0018-9162}
\urldef\tempurl%
\url{https://doi.org/10.1109/MC.2014.133}
\showDOI{\tempurl}


\bibitem[\protect\citeauthoryear{Schmidt and Roth}{Schmidt and Roth}{2014}]%
        {Schmidt2014ShrinkageFF}
\bibfield{author}{\bibinfo{person}{Uwe Schmidt} {and} \bibinfo{person}{Stefan
  Roth}.} \bibinfo{year}{2014}\natexlab{}.
\newblock \showarticletitle{Shrinkage Fields for Effective Image Restoration}.
\newblock \bibinfo{journal}{\emph{2014 IEEE Conference on Computer Vision and
  Pattern Recognition}} (\bibinfo{year}{2014}), \bibinfo{pages}{2774--2781}.
\newblock


\bibitem[\protect\citeauthoryear{Shi, Caballero, Huszar, Totz, Aitken, Bishop,
  Rueckert, and Wang}{Shi et~al\mbox{.}}{2016}]%
        {Shi2016pixelcnncvpr}
\bibfield{author}{\bibinfo{person}{Wenzhe Shi}, \bibinfo{person}{Jose
  Caballero}, \bibinfo{person}{Ferenc Huszar}, \bibinfo{person}{Johannes Totz},
  \bibinfo{person}{Andrew~P. Aitken}, \bibinfo{person}{Rob Bishop},
  \bibinfo{person}{Daniel Rueckert}, {and} \bibinfo{person}{Zehan Wang}.}
  \bibinfo{year}{2016}\natexlab{}.
\newblock \showarticletitle{Real-Time Single Image and Video Super-Resolution
  Using an Efficient Sub-Pixel Convolutional Neural Network}.
\newblock \bibinfo{journal}{\emph{2016 IEEE Conference on Computer Vision and
  Pattern Recognition (CVPR)}} (\bibinfo{year}{2016}),
  \bibinfo{pages}{1874--1883}.
\newblock


\bibitem[\protect\citeauthoryear{Timofte, Smet, and Gool}{Timofte
  et~al\mbox{.}}{2014}]%
        {Timofte2014aplusaccv}
\bibfield{author}{\bibinfo{person}{Radu Timofte}, \bibinfo{person}{Vincent~De
  Smet}, {and} \bibinfo{person}{Luc~Van Gool}.}
  \bibinfo{year}{2014}\natexlab{}.
\newblock \showarticletitle{A+: Adjusted Anchored Neighborhood Regression for
  Fast Super-Resolution}. In \bibinfo{booktitle}{\emph{ACCV}}.
\newblock


\bibitem[\protect\citeauthoryear{Wiener}{Wiener}{1964}]%
        {WienerFilter}
\bibfield{author}{\bibinfo{person}{Norbert Wiener}.}
  \bibinfo{year}{1964}\natexlab{}.
\newblock \bibinfo{booktitle}{\emph{Extrapolation, Interpolation, and Smoothing
  of Stationary Time Series}}.
\newblock \bibinfo{publisher}{The MIT Press}.
\newblock
\showISBNx{0262730057}


\bibitem[\protect\citeauthoryear{Yang, Wright, Huang, and Ma}{Yang
  et~al\mbox{.}}{2010}]%
        {Yang2010ImageSV}
\bibfield{author}{\bibinfo{person}{Jianchao Yang}, \bibinfo{person}{John~N.
  Wright}, \bibinfo{person}{Thomas~S. Huang}, {and} \bibinfo{person}{Yi Ma}.}
  \bibinfo{year}{2010}\natexlab{}.
\newblock \showarticletitle{Image Super-Resolution Via Sparse Representation}.
\newblock \bibinfo{journal}{\emph{IEEE Transactions on Image Processing}}
  \bibinfo{volume}{19} (\bibinfo{year}{2010}), \bibinfo{pages}{2861--2873}.
\newblock


\bibitem[\protect\citeauthoryear{Yue, Shen, Li, Yuan, Zhang, and Zhang}{Yue
  et~al\mbox{.}}{2016}]%
        {Yue2016ImageST}
\bibfield{author}{\bibinfo{person}{Linwei Yue}, \bibinfo{person}{Huanfeng
  Shen}, \bibinfo{person}{Jie Li}, \bibinfo{person}{Qiangqiang Yuan},
  \bibinfo{person}{Hongyan Zhang}, {and} \bibinfo{person}{Liangpei Zhang}.}
  \bibinfo{year}{2016}\natexlab{}.
\newblock \showarticletitle{Image super-resolution: The techniques,
  applications, and future}.
\newblock \bibinfo{journal}{\emph{Signal Processing}}  \bibinfo{volume}{128}
  (\bibinfo{year}{2016}), \bibinfo{pages}{389--408}.
\newblock


\bibitem[\protect\citeauthoryear{Zhang, Zuo, Chen, Meng, and Zhang}{Zhang
  et~al\mbox{.}}{2017a}]%
        {Zhang2017BeyondAG}
\bibfield{author}{\bibinfo{person}{Kai Zhang}, \bibinfo{person}{Wangmeng Zuo},
  \bibinfo{person}{Yunjin Chen}, \bibinfo{person}{Deyu Meng}, {and}
  \bibinfo{person}{Lei Zhang}.} \bibinfo{year}{2017}\natexlab{a}.
\newblock \showarticletitle{Beyond a Gaussian Denoiser: Residual Learning of
  Deep CNN for Image Denoising}.
\newblock \bibinfo{journal}{\emph{IEEE Transactions on Image Processing}}
  \bibinfo{volume}{26} (\bibinfo{year}{2017}), \bibinfo{pages}{3142--3155}.
\newblock


\bibitem[\protect\citeauthoryear{Zhang, Zuo, Gu, and Zhang}{Zhang
  et~al\mbox{.}}{2017b}]%
        {kai2017ircnncvpr}
\bibfield{author}{\bibinfo{person}{Kai Zhang}, \bibinfo{person}{Wangmeng Zuo},
  \bibinfo{person}{Shuhang Gu}, {and} \bibinfo{person}{Lei Zhang}.}
  \bibinfo{year}{2017}\natexlab{b}.
\newblock \showarticletitle{Learning Deep CNN Denoiser Prior for Image
  Restoration}.
\newblock \bibinfo{journal}{\emph{2017 IEEE Conference on Computer Vision and
  Pattern Recognition (CVPR)}} (\bibinfo{year}{2017}),
  \bibinfo{pages}{2808--2817}.
\newblock


\bibitem[\protect\citeauthoryear{Zhang, Zuo, and Zhang}{Zhang
  et~al\mbox{.}}{2018}]%
        {kai2018srmdcvpr}
\bibfield{author}{\bibinfo{person}{Kai Zhang}, \bibinfo{person}{Wangmeng Zuo},
  {and} \bibinfo{person}{Lei Zhang}.} \bibinfo{year}{2018}\natexlab{}.
\newblock \showarticletitle{Learning a Single Convolutional Super-Resolution
  Network for Multiple Degradations}.
\newblock \bibinfo{journal}{\emph{2018 IEEE/CVF Conference on Computer Vision
  and Pattern Recognition}} (\bibinfo{year}{2018}),
  \bibinfo{pages}{3262--3271}.
\newblock


\end{thebibliography}
\end{document}